\documentclass[aps, twocolumn, superscriptaddress]{revtex4-1}

\usepackage{graphicx}
\usepackage{amsmath}
\usepackage{dsfont}
\usepackage{amssymb}
\usepackage{physics}
\usepackage{hyperref}
\usepackage{ulem}
\hypersetup{colorlinks=true,
	    final=true,
	    linkcolor=blue,
	    citecolor=blue,
	    filecolor=blue,
	    urlcolor=blue,}

\newcommand{\rta}{\rightarrow}

\newcommand{\ra}{\rangle}
\newcommand{\la}{\langle}

\newcommand{\beq}{\begin{equation}}
\newcommand{\eeq}{\end{equation}}

\newcommand{\vbar}{\bar{v}}

\begin{document}

\title{Drude's lesser known error of a factor of two and Lorentz's correction}
\author{Navinder Singh}
\email{navinder.phy@gmail.com}
\affiliation{Theoretical Physics Division, Physical Research Laboratory, Ahmedabad, India. PIN: 380009.}

\begin{abstract}
As is well known, Paul Drude put forward the very first quantitative theory of electrical conduction in metals in 1900. He could successfully account for the Wiedemann-Franz law which states that the ratio of thermal to electrical conductivity divided by temperature is a constant called the Lorenz number.  As it turns out, in Drude's derivation, there is a lucky cancellation of two errors. Drude's under-estimate (by an order of 100) of the  value of square of the average electron velocity compensated his over-estimate of the electronic heat capacity  (by the same order of 100).  This compensation or cancellation of two errors lead to a value of the Lorenz number very close to its experimental value. This is well known. There is another error of a factor of two which Drude made when he calculated two different relaxation times for heat conductivity and electrical conductivity. In this article we highlight how and why this error occurred in Drude's derivation and how it was removed 5 years later (that is in 1905) by Hendrik Lorentz when he used the Boltzmann equation and a single relaxation time. This article is of pedagogical value and may be useful to undergraduate/graduate  students learning solid state physics.
\end{abstract}

\maketitle

\section{Introduction}

Drude's theory for electrical conduction in metals has a very special place in the learning of the solid state physics. It is often the very first topic that students learn while learning the elementary solid state physics\cite{ash}.  There are subtleties involved in the definition and use of the terms like mean or average acceleration time, mean time between two successive collisions, relaxation time etc. In particular, subtleties involved in defining the mean acceleration time and mean time between two collisions lead to an error of a factor of two in Drude's calculation of the Lorenz number. These issues are sorted out here. 

Paul Drude at the turn of the last century (in 1900) put forward the very first quantitative theory of electrical conduction in metals. The subject of classical kinetic theory as developed by Maxwell and Boltzmann was in place when Drude advanced his theory. Electron was discovered by J. J. Thomson in 1897, three years prior to the Drude theory. So Paul Drude had enough motivation to assume that charge carriers in metals must be electrons and that these should be treated as classical particles much like gas molecules so that kinetic theory approaches of Maxwell and Boltzmann can be applied. For the mechanism of resistivity he assumed the impediment of the motion of electrons when they are forced to flow in a particular direction under the action of an external voltage difference. The impediment of motion occurs when they suffer collisions with static ion cores (as assumed by Drude) which randomizes their velocities and this is the basic mechanism of resistivity in his theory.

As is well known, in developing his concrete mathematical model Drude relied on the following basic assumptions:

\begin{enumerate}
\item Electrons collide with immobile ion cores (electron-electron scattering is neglected).
\item Each collision randomizes electron's velocity.
\item And an average time between two successive collisions can be defined.
\end{enumerate}

In next two sections we sketch two different derivations of the Drude formula. The derivation no. 1 is based on a "snapshot" method, and the derivation no. 2 is based on probability method. Both are internally consistent. However, only the second one corresponds to the way relaxation time is defined in the relaxation time approximation solution of the Boltzmann equation. In the processes of derivations the subtleties involved will be discussed and clarified. 

\section{Derivation 1 ("snapshot" method)}

Let us assume that a metallic sample is biased with an external potential difference and the impressed electric field is given by $E_x$ (along the x-direction). The induced current density is given by:

\beq
J_x = -n e \vbar_x.
\eeq 

\begin{figure}[h!]
    \centering
    \includegraphics[width=0.5\columnwidth]{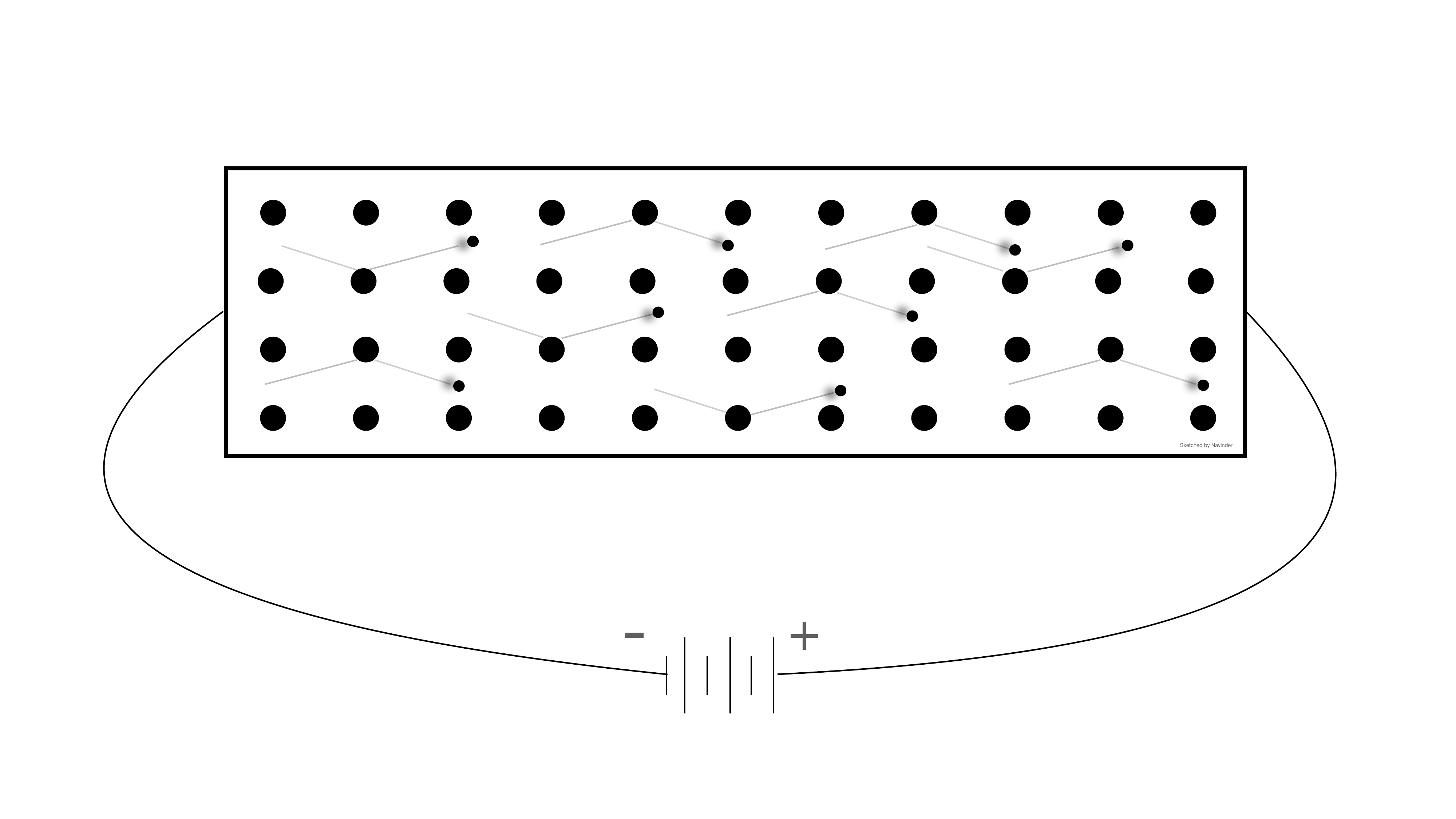}
    \caption{The Drude model.}
    \label{f1}
\end{figure}

Here $n$ is the number of electrons per unit volume in the sample, $\vbar_x$ is the average drift velocity of electrons in the $x-$direction. In the "snapshot" method we consider a particular instant of time and look at the history of motion of electrons (just imagine that we are making a movie of electrons and a photograph is clicked at a given instant of time. In this hypothetical photograph we notice that some electrons have just collided with ion cores, some are about to collide, and many are in their flight on their trajectories). To be specific let us focus our attention on an electron, say $i^{th}$ electron. Assume that it had its last collision some time back and it was accelerating for $t_i$ seconds during which x-component of its velocity changed from its initial value of $u_{0,x}^i$ to a final value $u_x^i$ which is given by:

\beq
u_x^i = u_{x,0}^i - \frac{e E_x}{m} t_i.
\eeq

Similarly, another electron, say $j^{th}$, was accelerating for a time $t_j$ after its last collision. x-component of it's velocity at the "snapshot" time will be:

\beq
u_x^j = u_{x,0}^j - \frac{e E_x}{m} t_j.
\eeq

Similar equations can be written for all the electrons in the sample. We add all such equations and divide by the total number of electrons in the sample ($N$, say) we get:

\beq
\frac{1}{N}\sum_{i=1}^N u_x^i = \frac{1}{N}\sum_{i=1}^N u_{x,0}^i - \frac{e E_x}{m}  \frac{1}{N}\sum_{i=1}^N t_i.
\eeq
The term on the left hand side is an average drift velocity ($\vbar_x$) along the $x$ direction. First term on the right hand side vanishes due to the assumption 2 above.  Part of the second term on the right hand side ($\frac{1}{N}\sum_{i=1}^N t_i$) is an average time for which electrons "feel" acceleration, or, an average time for acceleration ($\tau_a$).  From equations (1) and (4) and implementing  the above, we get

\beq
J_x = \frac{ne^2\tau_a}{m} E_x.
\eeq

On comparing it with Ohm's law $J_x = \sigma_{xx} E_x$ we find that 

\beq
\sigma =\sigma_{xx} = \frac{n e^2\tau}{m}.
\eeq

This is the famous Drude's law for electrical conductivity. The point to be noted is that $\tau_a$ is an average time for which electrons suffer {\it acceleration}. This is specific to this "snapshot" method, and it is given by

\beq
\tau_a =  \frac{t_i + t_j + t_k+......}{N} = \frac{1}{N}\sum_{i=1}^N t_i.
\eeq

Next, let us calculate the thermal conductivity within this method. If the average time for acceleration is $\tau_a$, then the average time between two successive collisions will be $2\tau_a$. The proof goes like this: Let us again look at the motion of $i^{th}$ electron. It was accelerating for time $t_i$, let us assume that it remains accelerating for another $t_i'$ seconds just before it suffers collision.  Therefore, from the instant  considered (our "snapshot" time) it accelerates for another $t_i'$ seconds. The $j^{th}$ electron accelerates for another $t_j'$ seconds (from our snapshot instant) before it suffers collision. Similarly, for the $k^{th}$ electron further acceleration time is $t_k'$ etc. The average time for which all the electrons will suffer acceleration (from the "snapshot" instant under consideration) is given by

\beq
\tau_a = \frac{t_i' + t_j' + t_k'+......}{N}.
\eeq

By symmetry of the time evolution, and by the law of large numbers this time scale is also $\tau_a$ (This can be different $\tau_a'$ only if a small number of electrons are involved. But here, we are dealing with very large number of electrons, of the order of Avogadro number). Thus the average total time between two successive collisions will be $2\tau_a$. We will compare and contrast this derivation with another derivation (derivation 2 in section III).

 Coming back to Drude's work, for the calculation of thermal conductivity, he directly used the expression for thermal conductivity previously obtained from kinetic theory methods:

\beq
\kappa =\frac{1}{3}c_v \vbar l.
\eeq 

Here $c_v$ is the heat capacity per unit volume for electrons, $\vbar$ is an average speed, and $l$ is the mean free path length (average distance travelled by an electron between two successive collisions). For $l$ he set $l= \vbar  2 \tau_a$\cite{fn1}. For heat capacity, Drude used the classical expression $c_v = \frac{3}{2} n k_B$ (treating electrons as classical free particles with 3 degrees of freedom). To calculate $\vbar$ he used equipartition theorem result: $\frac{1}{2} m \vbar^2 =\frac{3}{2}k_B T$. With all these, updated formula for thermal conductivity takes the form:

\beq
\kappa = \frac{3 n k_B^2 T \tau_a}{m}.
\eeq

The ratio $\frac{\kappa}{\sigma T}$ is called the Lorenz number (not to be confused with Lorentz). Its value from Drude theory (equations (6) and (10)) is:

\beq
L = \frac{\kappa}{\sigma T} = 3 \left(\frac{k_B}{e}\right)^2.
\eeq

Numerical value of it is $2.23 \times 10^{-8} V^2 K^{-2}$ which is much closer to the typical experimental value  $2.4 \times 10^{-8} V^2 K^{-2}$ for most of the metals at room temperature. This was the major triumph of the Drude theory. However, when correct calculation (refer to next paragraph) is used the Lorenz number turns out to be $L = \frac{\kappa}{\sigma T} = \frac{3}{2} \left(\frac{k_B}{e}\right)^2$ (half the value obtained by Drude). That is, if  $l= \vbar \tau_a$ is used in the thermal conductivity formula, then the correct expression (half of the value of the above equation) is obtained. Now the Lorentz number reads $1.11 \times 10^{-8} V^2 K^{-2}$ which is in worse agreement with experimental value.

Drude's error of a factor of two occurred because he used the thermal conductivity formula directly without attempting its derivation consistently within the assumed setting\cite{dru}!

To observe that the factor of two error can be immediately removed if the thermal conductivity calculation is done in the same setting ('snapshot' method) as the electrical conductivity calculation is done.  Let us consider the following experimental situation. A thin metallic rod is connected at its ends with two thermal reservoirs, one hotter and the other colder. Let $J_q^x$ be the heat current  density (Joules per second per unit area) flowing along the x-direction (along the axis of the rod). Let us assume that a uniform temperature gradient ($dT(x)/dx$) is set up. Imagine a fictitious plane perpendicular to the axis of the rod at a distance $x$ from the hotter end. We want to compute the heat current density flowing through this fictitious plane. As above, electrons coming from the hotter side would have no collision for $\tau_a$ seconds on the average. Thus they are having average thermal energy of $U(T(x-\vbar \tau_a))$. Total $n/2$ electrons are traveling from hotter to colder end, and $n/2$ in the opposite direction (zero net charge transport).  The average thermal energy transported from hotter end to colder end through that fictitious plane will be $\frac{n}{2} \vbar_x U(x-\vbar_x \tau_a)$. Actually, we should also consider spatial variation in mean electron velocity $\vbar_x$ such as average of $v_x(x-\vbar_x\tau_a)$. But then, Taylor expansions that we are going to do under the valid approximation ($\vbar_x \tau_a<<x$ ($x$ is a macroscopic length scale) and functions $U$ and $T$ are smooth functions) will lead to additional terms in the Fourier's law $J_q =- \kappa \nabla T$. We assume Fourier's law to be valid thus consider only one single average velocity $\vbar_x$. Similarly, the heat current in the opposite direction will be $-\frac{n}{2} \vbar_x U(x+\vbar_x \tau_a)$.  Adding both the terms gives the net heat current flowing from hotter end to colder end:

\beq
J_q^x = \frac{n}{2} \vbar_x U(x-\vbar_x \tau_a) - \frac{n}{2} \vbar_x U(x+\vbar_x \tau_a) \simeq  - n \vbar_x^2 \tau_a \frac{dU}{dT}\frac{dT}{dx}.\nonumber\
\eeq

$n \frac{dU}{dT}$ is the total electronic heat capacity ($c_v$) per unit volume. By considering the isotopy of the space ($\bar{v_x^2} = \bar{v_y^2} = \bar{v_z^2} = \frac{1}{3}\vbar^2$), we can quickly generalize the above result to the 3D case:

\beq
J_q =- \kappa \nabla T,~~~~~J_q = -\frac{1}{3}\vbar^2\tau_a c_v \nabla T.
\eeq 

Yielding the conductivity:

\beq
\kappa = \frac{1}{3}\vbar^2\tau_a c_v 
\eeq

When the ratio of the above equation is taken with that for conductivity, we get

\beq
L = \frac{\kappa}{\sigma T} = \frac{3}{2} \left(\frac{k_B}{e}\right)^2.
\eeq

Which is half of the value obtained by Drude (compare equations (11) and (14)). But if the formula  $\kappa = \frac{1}{3} l \vbar c_v$ is directly used and the value of $l$ is set to $2 \vbar\tau_a$ one gets  the wrong result (as Drude obtained). Thus the error of a factor of two occurred when Drude directly used the formula of thermal conductivity without attempting its full derivation consistently within the assumed setting\cite{dru}. Important point to be noted is that Drude did not commit any error (within snapshot method)  when $\tau_a$  is defined as the mean acceleration time and $2\tau_a$ as a mean time between to successive collisions.

We will go through another derivation of the Drude formula which actually corresponds to the definition of $\tau$ as the relaxation time or an average time for one collision.  This corresponds to the relaxation time used in the case of Boltzmann equation under relaxation time approximation  which Lorentz used in 1905\cite{}.

\section{Derivation 2 (Probability method)}

Define $\tau$ in the following way: The probability that an electron undergoes a collision in a time interval from $t$ to $t+dt$ is given as $\frac{dt}{\tau}$. Probability that it does not collide in that interval is $1-\frac{dt}{\tau}$. Defined this way, $\tau$ is an average time scale for {\it one} collision of an electron.  With this definition of $\tau$ it turns out that an average time for all electrons back to their last collision or up to their next collision is $\tau$. But it does not mean that average time between two successive collisions is $2\tau$. In fact, it means that the average time between two successive collisions is also $\tau$. This seems counter-intuitive (compare and contrast with "snapshot" method) but this is true within the way $\tau$ is defined now. This is the essence of the problem no. 1 in the famous textbook\cite{ash}. Probability that it had no collision during the preceding $t$ seconds is given as: $(1-\frac{\Delta t}{\tau}) (1-\frac{\Delta t}{\tau}) ......(N~ factors) $ where the interval $t$ is divided into large number of sub-intervals of width $\Delta t$ and $t = N \Delta t$, and where $N$ is a large number.  In the limit $N \rta \infty, ~and ~\Delta t \rta 0$, the product of $N$ factors gives $p(t) = e^{-t/\tau}$.  Probability that it had no collision during the preceding $t$ seconds is $e^{-t/\tau}$. Thus for a large $t$ (much larger than $\tau$) probability that it had no collision is essentially zero. Similarly, probability that it will have no collision in the next $t$ seconds is $p(t) = e^{-t/\tau}$.

Average time or mean time for all the electrons can be calculated from the probability $p(t)$. Average time for all electrons back to their last collision or up to their next collision is $\la t\ra = \frac{\int_0^\infty t p(t) dt}{\int_0^\infty p(t) dt} =\tau$. To show that average time between two successive collisions is $\tau$ not $2\tau$,  define another probability distribution $p_2(t_2)$ which consists of a product (convolution) of two probabilities:

\beq
p_2(t_2)\frac{dt_2}{\tau} = \int_0^{t_2} p(t_1) \frac{dt_1}{\tau} p(t_2-t_1)\frac{dt_2}{\tau},
\eeq
or,
\beq
p_2(t_2) dt_2 = \frac{t_2}{\tau} e^{-t_2/\tau}dt_2.
\eeq

This means that we have a time interval $0$ to $t_2$ and we divide it into two parts, first part is from $0$ to $t_1$ and the second part is from $t_1$ to $t_2$. We demand that there are no collisions from $0$ to $t_1$ and  then there is a collision at $t_1$ given as $\frac{dt_1}{\tau}$. This leads to the first factor $p(t_1)\frac{dt_1}{\tau}$ in the integrand. We again demand that there are no collisions from $t_1$ to $t_2$ and then there is a collision at time $t_2$ given as $\frac{dt_2}{\tau}$.  This corresponds to the second factor $p(t_2-t_1)\frac{dt_2}{\tau}$ in the integrand. The average time for this whole process is $\frac{\int_0^\infty t_2 p_2(t_2)dt_2}{\int_0^\infty p_2(t_2)dt_2} = 2\tau$. But we are interested in the average value of the second part of the process that is between $t_1$ and $t_2$. This part of the trajectory is between two successive collisions. Average value of it will be an average time between two successive collisions. This is given by

\beq
 \frac{\int_0^\infty t_2 p_2(t_2)dt_2}{\int_0^\infty p_2(t_2)dt_2} - \frac{\int_0^\infty t p(t) dt}{\int_0^\infty p(t) dt} = \tau.
\eeq

This proves the point that average time between two successive collisions is also $\tau$. But in the snapshot method  (derivation no. 1) it is $2\tau_a$. This is so because it is defined  in the specific way considering acceleration of electrons up to  specific time (snapshot time). Within this "probability method" both formulas (for electrical conductivity and thermal conductivity) can be easily derived as before or as in\cite{ash} and Lorenz number has the correct value $\frac{3}{2}(\frac{k_B}{e})^2$ (within the classical model)\cite{}.

At the outset it appears to be a minor point but it has been a point of confusion in the literature. For example, in the text book by A. H. Wilson this oversight has occured\cite{wil} and the Lorenz number is twice the actual value.   Also this point is glossed over in another textbook\cite{kub}. Only the books, such as\cite{ash,cham}, present it correctly but the reason why this occurs is not properly addressed.  In author's book\cite{nav} also this oversight has occurred and will be corrected in the new edition of it.

\section{Summary}

In conclusion, Drude's error occurred because he directly used the expression of the thermal conductivity without attempting its derivation within the assumed setting. Had he used the probabilistic method (same mean time scale for acceleration and same mean time between two collisions) he would have obtained the correct result even without attempting the derivation of the thermal conductivity formula.  But this is a part of the history. Both methods presented above are internally consistent and leads to the correct value of the Lorenz number. However, the second method is equivalent to the relaxation time approximation in the Boltzmann equation and is the standard one.

\begin{acknowledgments}
Author thanks Paramita Dutta for carefully reading the manuscript and suggestions.
\end{acknowledgments}

%---------------------------------------


\begin{thebibliography}{99}
\bibitem{fn1}  Actually, Drude set the average time between two collisions as $\tau_a$, thus for the average time of acceleration he took $\frac{\tau_a}{2}$, and the expression obtained by him for conductivity reads $\sigma = \frac{ne^2\tau_a}{2 m}$. That is half of the standard value.
\end{thebibliography}
\end{document}